\documentclass[aps,prl,twocolumn,showpacs,superscriptaddress,floatfix]{revtex4}

\usepackage{amsmath,amsfonts,amssymb,graphics,graphicx,epsfig,color,times}
\usepackage{graphicx}
\usepackage{epsfig}
\usepackage[english]{babel}

\newcommand{\be}{\begin{equation}}
\newcommand{\bea}{\begin{eqnarray}}
\newcommand{\eea}{\end{eqnarray}}
\newcommand{\ee}{\end{equation}}

\newcommand{\ket}[1]{\mbox{$| #1 \rangle$}}

\newcommand{\HH}{\mathcal{H}}

\newcommand{\RT}{\mathcal{R}}
\newcommand{\ST}{\mathcal{S}}
\newcommand{\TT}{\mathcal{T}}

\def\one{\ensuremath{\hbox{$\mathrm I$\kern-.6em$\mathrm 1$}}}
\def\tr{ \mbox{tr}}
\def\qed{\leavevmode\unskip\penalty9999 \hbox{}\nobreak\hfill
     \quad\hbox{\leavevmode  \hbox to.77778em{%
               \hfil\vrule   \vbox to.675em%
               {\hrule width.6em\vfil\hrule}\vrule\hfil}}
     \par\vskip3pt}
\newcommand{\beaa}{\begin{eqnarray*}}
\newcommand{\eeaa}{\end{eqnarray*}}
\newcommand{\bma}{\begin{subequations}}
\newcommand{\ema}{\end{subequations}}

\def\one{{\bf 1}}

\def\noxrightarrow[#1]{\dodoublegroupempty\dodoxrightarrow{#1}}
\def\noxleftarrow [#1]{\dodoublegroupempty\dodoxleftarrow {#1}}
\def\dodoxrightarrow#1#2{\mathrel{{\domthxarr0359\rightarrowfill{#1}{#2}}}}
\def\dodoxleftarrow#1#2{\mathrel{{\domthxarr3095\leftarrowfill{#1}{#2}}}}

\begin{document}

\date{\today}

\title{\bf Quantum computation, quantum state engineering,
and quantum phase transitions\\ driven by dissipation}

\author{Frank Verstraete}
\affiliation{Fakult\"at f\"ur Physik, Universit\"at Wien,
Boltzmanngasse 5, A-1090 Wien, Austria.}
\author{Michael M. Wolf}
\author{J. Ignacio Cirac}
\affiliation{Max-Planck-Institut f\"ur Quantenoptik,
Hans-Kopfermann-Str. 1, D-85748 Garching, Germany.}

\begin{abstract}
We investigate the computational power of creating steady-states
of quantum dissipative systems whose evolution is governed by
time-independent and local couplings to a memoryless environment.
We show that such a model allows for efficient universal quantum
computation with the result of the computation encoded in the
steady state. Due to the purely dissipative nature of the process,
this way of doing quantum computation exhibits some inherent
robustness and defies some of the DiVincenzo criteria for quantum
computation. We show that there is a natural class of problems
that can be solved with such a model---the preparation of ground
states of frustration free quantum Hamiltonians. This allows for
robust and efficient creation of exotic states that exhibit
features like topological quantum order and the creation of PEPS
and it proves the existence of novel dissipative phase
transitions. In particular the latter can in principle be verified
experimentally with present day technology such as with optical
lattices.

\end{abstract}


\maketitle

The strongest adversary in quantum information science is
decoherence, which arises due to the coupling of a system with its
environment. The induced dissipation tends to destroy and wash out
the interesting quantum effects which give rise to the power of
quantum computation, cryptography, and simulation. While this is
certainly true for many forms of dissipation, we show here that
dissipation can also have exactly the opposite effect: it can be a
full--fledged resource for universal quantum computation and
quantum state engineering without any coherent dynamics needed to
complement it.

We consider a quantum system composed of $N$ particles (e.g.
qubits) interacting with local environments giving rise to
memoryless and time--independent dissipation processes. We will
show first how to design the interactions with the environment to
implement universal quantum computation. This new method, which we
refer to as dissipative quantum computation (DQC), defies some of
the standard criteria for quantum computation since it requires
neither state preparation, nor unitary dynamics \cite{DiVincenzo}.
However, it is nevertheless as powerful as standard quantum
computation. Then we will show that dissipation can be engineered
\cite{poyatos} to prepare all ground states of frustration free
Hamiltonians. Those include matrix product states (MPS) \cite{MPS}
and projected entangled--pair states (PEPS) \cite{PEPS}, like
graph states \cite{graphstates} and Kitaev \cite{toric} and
Levin--Wenn \cite{Levin-Wen} topological codes. Both DQC and
dissipative state engineering (DSE) are robust in the sense that,
given the dissipative nature of the process, the system is driven
towards its steady state independent of the initial state and
hence of eventual perturbations along the way.

Apart from novel ways of performing quantum computation or state
engineering, our results imply that quantum phase transitions
\cite{Sachdev} may be driven by dissipation alone. That is, the
physical properties of the steady state of our system may change
abruptly in the thermodynamical limit when we slightly modify the
parameters characterizing the dissipative dynamics \cite{note1}.
This immediately follows from the existence of frustration free
quantum Hamiltonians that exhibit quantum phase transitions in one
\cite{MPSphase} and two dimensions \cite{PEPSphase}.

In this work we will concentrate first on DQC, showing how given
any quantum circuit one can construct a master equation whose
steady state is unique, encodes the outcome of the circuit, and is
reached in polynomical time (with respect to the one corresponding
to the circuit). Then we will show how to construct dissipative
processes which drive the system to the ground state of any
frustration free Hamiltonian. We will prove that MPS and certain
kinds of PEPS can be efficiently prepared using this method, with
details given in the appendix. In this paper we will not consider
specific physical setups where our ideas can be implemented.
Nevertheless, the appendix will provide a universal way of
engineering the master equations required for DQC and DSE, which
can be easily adapted to current experiments based on, e.g. atoms
in optical lattices \cite{reviewopticallattices} or trapped ions
\cite{reviewtrappedions}. Thus, we expect that our predictions and
in particular the existence of quantum phase transitions driven by
dissipation may be experimentally tested in the near future.

Let us start  with DQC by considering $N$ qubits in a line and a
quantum circuit specified by a sequence of nearest--neighbor qubit
operations $\{U_t\}_{t=1}^T$
. We define $|\psi_t\rangle:= U_{t}U_{t-1}...U_1|0\rangle_1
\otimes ... |0\rangle_N$, so that $|\psi_T\rangle$ is the final
state after the computation. Our goal is to find a master equation
$\dot\rho={\cal L}(\rho)$ with Liouvillian in Lindblad form
\cite{Lindblad}
 \be
 \label{Lindblad}
 {\cal L}(\rho)=\sum_k L_k\rho L_k^\dagger -\frac{1}{2}\left\{L_k^\dagger
 L_k,\rho\right\}_+,
 \ee
where the $L_k$ act locally and has a steady state, $\rho_0$: (i)
which is unique; (ii) that can be reached in a time poly$(T)$;
(iii) such that $\psi_T$ can be extracted from it in a time
poly$(T)$. As in Feynman's construction of a quantum simulator
\cite{Feynman}, we consider another auxiliary register with states
$\{|t\rangle\}_{t=0}^T$, which will represent the time. We choose
the Lindblad operators
 \bea
 L_i&=&|0\rangle_i\langle 1|\otimes |0\rangle_t\langle 0|\\
 L_t&=& U_t\otimes|t+1\rangle\langle t|+U_t^\dagger\otimes|t\rangle\langle
 t+1|,
 \eea
where $i=1,\ldots, N$ and $t=0,\ldots,T$. It is clear that the
$L$'s act locally except for the interaction with the extra
register, which can be made local as well. Furthermore,
 \be
 \rho_0 = \frac{1}{T+1}\sum_t |\psi_t\rangle\langle\psi_t|\otimes |t\rangle\langle
 t|.\label{fixed}
 \ee
is a steady state, i.e. ${\cal L}(\rho_0)=0$. Given such a state,
the result of the actual quantum computation can be read out with
 probability $1/T$ by measuring the time register. In the
appendix we show that $\rho_0$ is the unique steady state and that
the Liouvillian has a spectral gap
$\Delta =\pi^2/(2T+3)^2$. This means indeed that the steady state
will be reached in polynomial time in $T$. Note that this gap is
independent of $N$ as well as on the actual quantum computation
which is performed (i.e. independent of the $U_t$). It is also
shown that the same gap is retained if the clock register is
encoded in the unary way proposed by Kitaev \cite{KitaevQMA},
making the Lindblad operators strictly local. A sketch of the
proof is as follows: first, we do a similarity transformation on
${\cal L}$ that replaces all gates $U_i$ with the identity gates,
showing that its spectrum is independent of the actual quantum
computation. Second, another similarity transformation is done
that makes ${\cal L}$ hermitian and block-diagonal. Each block can
then be diagonalized exactly leading to the claimed gap.

In some sense, the present formalism can be seen as a robust way
of doing adiabatic quantum computation \cite{adiabatic} (errors do
not accumulate and the path does not have to be engineered
carefully) and implementing quantum random walks
\cite{randomwalks}, and it might therefore be easier to tackle
interesting open questions, such as the quantum PCP theorem, in
this setting \cite{PCP}. Also, it seems that the dissipative way
of preparing ground states is more natural than to use adiabatic
time evolution, as nature itself prepares them by cooling.

Let us now turn to DSE and consider again a quantum system with
$N$ particles on a lattice in any dimension. We are interested in
ground states $\Psi$, of Hamiltonians
 \be
 \label{frusfree}
 H=\sum_{\lambda} H_\lambda,
 \ee
which are frustration free, meaning that $\Psi$ minimizes the
energy of each $H_\lambda$ individually, and local in the sense
that $H_\lambda$ acts non-trivially only on a small set
$\lambda\subset\{1,\ldots, N\}$ of sites (e.g., nearest neighbors)
\cite{note3}. We can assume the $H_\lambda$'s to be projectors and
we will denote the orthogonal projectors by
$P_\lambda=\one-H_\lambda$. States $\Psi$ of the considered form
are, e.g., all PEPS (including MPS and stabilizer
states\cite{stabilizer}).

We will consider discrete time evolution generated by a trace
preserving completely positive map (cp-map) instead of a master
equation. These two approaches are basically equivalent
\cite{dividing} as every local cp-map ${\cal T}$ can be associated
to a local Liouvillian via ${\cal L}(\rho)=N[{\cal T}(\rho)-\rho]$,
which leads to the same fixed points and a gap that is $N$ times
larger\cite{note4}. We choose cp-maps of the form
 \be\label{eq:T}
 {\cal T}(\rho)=\sum_\lambda p_\lambda\left[ P_\lambda\rho P_\lambda +
 \frac1m\sum_{i=1}^m U_{\lambda,i} H_\lambda \rho H_\lambda
 U _{\lambda,i}^\dagger\right],
 \ee
where the $p_\lambda$'s are probabilities and
$U_{\lambda,1},\ldots,U_{\lambda,m}$ is a set of unitaries acting
non-trivially only within region $\lambda$. They effectively
rotate part of the high-energy space (with support of $H_\lambda$)
to the zero-energy space, so that $\tr[{\cal T}(\rho)\Psi]\geq
\tr[\rho\Psi]$ increases. As for Liouvillians (\ref{Lindblad}) we
could similarly take $L_{\lambda,i}=U_i H_\lambda$, or the ones
associated to the cp-map.

We show now that for every frustration-free Hamiltonian the cp-map
in Eq.(\ref{eq:T}) converges to the ground state space if we
choose the unitaries $U_{\lambda,i}$ to be completely
depolarizing, i.e., ${\cal T}(\rho)\propto \sum_\lambda P_\lambda
\rho P_\lambda
+\one_\lambda\otimes\tr_\lambda[H_\lambda\rho]/\tr[\one_\lambda]
$. For ease of notation we will explain the proof for the case of
a one-dimensional ring with nearest-neighbor interactions labelled
by the first site $\lambda=1,\ldots,N$. Assume $\rho$ is such that
its expectation value with respect to the projector $\Psi$ onto
the ground state space of $H$ is non-increasing under applications
of $\cal T$, i.e., in particular $\tr[\rho\Psi]=\tr[{\cal
T}^N(\rho)\Psi]$. Expressing this in the Heisenberg picture in
which ${\cal T}^*(\Psi)=\Psi +  \sum_\lambda H_\lambda\tr_\lambda
(\Psi)/(d^2 N)$ we get
 \bea
 \tr[\rho\Psi]&\geq& \tr[\rho\Psi]+ \frac1{(d^2N)^{N}}\tr\left[\rho
 \sum_{\mu=1}^N \prod_{\lambda=1}^N
 \left(H_{\lambda+\mu} \tr_{\lambda+\mu}\right)(\Psi)\right]\nonumber \\
 &\geq& \tr[\rho\Psi]+\frac{\nu^N}{(d^2N)^{N}}\tr[\rho H],
 \eea
where the first inequality comes from discarding (positive) terms
in the sum and the second one is due to bounding all partial
traces of $H_\lambda$ from below by the respective smallest
eigenvalue $\nu$. Note that the latter is strictly positive unless
$H$ has a product state as ground state (in which case the
statement becomes trivial). Hence, we must have $\tr[\rho H]=0$,
i.e., $\rho$ is  a ground state of $H$. It is easily seen that the
same argumentation goes through for more general interactions on
arbitrary lattices.

The above procedure implies the existence of quantum phase
transitions driven by dissipation. By changing the parameters in
the cp--map (or the master equation) one can obtain that some
physical properties of the steady state abruptly change, in as
much the same way as in Refs. \cite{MPSphase,PEPSphase}.

We have shown that it is possible to engineer dissipative
processes which prepare ground states of frustration free
Hamiltonians in steady state. However, in the above proof the time
 for this preparation scales as $N^N$, which may be an issue
for experiment with large number of particles. In the following we
give much more efficient method for certain classes of frustration
free Hamiltonians: commuting Hamiltonians and MPS.

We consider first frustration free Hamiltonians for which
$[H_\lambda,H_\mu]=0$ and show that the corresponding ground
states can be prepared in a time that only scales polynomially
with the number of particles. The corresponding set of ground
states contains important families, like stabilizer states (e.g.
cluster states and topological codes), or certain kinds of PEPS.
Note that there was no known way of efficient preparation for the
latter.

Loosely speaking there are two classes of Hamiltonians of this
type: (i) Hamiltonians for which all excitations can be locally
annihilated. In this case the time of convergence scales as
$\tau={\cal O}(\log N)$. (ii) Interactions where excitations have
to be moved along the lattice before they can annihilate and
$\tau={\cal O}(N\log{N})$.

In order to see how the first case can occur consider
Eq.(\ref{eq:T}) and note that it can be interpreted as randomly
choosing a region $\lambda$ (according to $p_\lambda$ which we may
set equal to $1/N$), then measuring $P_\lambda$ and applying a
correction according the $U$'s if the outcome was negative. Assume
now that when iterating $\cal T$ the correction on $\lambda$ does
not change the outcome of previous measurements on neighboring
regions since \be \label{eq:comU} \forall \lambda\neq\lambda':\ \
[U_{\lambda,i},H_{\lambda'}]=0.\ee In fact, this can always be
achieved by regrouping the regions into larger ones having an
interior $I(\lambda)\subset\lambda$ on which only $H_\lambda$ acts
non-trivially \footnote{For a 1D system with nearest-neighbor
interaction we can for instance choose
$H_{(1,2,3)}'=H_{(1,2)}+H_{(2,3)}$ with site 2 as interior.} and
letting the $U_{\lambda,i}$ solely act on $I(\lambda)$. Denote by
$q$ the largest probability for obtaining twice a negative
measurement outcome on the same region $\lambda$. The energy
$\tr[H{\cal T}^M(\rho)]$ after $M$ applications of $\cal T$
decreases then as $N(1-(1-q)/N)^M$ such that it takes ${\cal
O}\big((N\log N)/(1-q)\big)$ steps to converge to a ground state.
The relaxation time of the corresponding Liouvillian is thus
$\tau={\cal O}\big(\log N^{\frac1{1-q}}\big)$. Clearly, this is
only a reasonable bound if $q<1$, a condition possibly
incompatible with Eq.(\ref{eq:comU}).

Note that for all stabilizer states we can achieve $q=0$, since
there exists always a local unitary (acting on a single qubit) so
that $H_\lambda U_\lambda H_\lambda=0$. A class of stabilizer
states where this is compatible with Eq.(\ref{eq:comU}) are the
so-called \emph{graph states} \cite{graphstates}. In this case
$\lambda$ labels (with some abuse of notation) a vertex of a graph
and
$H_\lambda=(\one-\sigma_x^{(\lambda)}\prod_{(\lambda,\mu)\in\cal
E} \sigma_z^{(\mu)})/2$ where $\sigma^{(\lambda)}$ is a Pauli
operator acting on site $\lambda$ and $\cal E$ is the set of edges
of the graph. Obviously, $U_\lambda=\sigma_z^{(\lambda)}$ does the
job. In this special case we can get even faster convergence when
using the Liouvillian \be\label{eq:graphL}{\cal
L}(\rho)=\Big(\sum_\lambda U_\lambda H_\lambda\rho H_\lambda
U_\lambda^\dagger\Big)-\frac12\big\{H,\rho\big\}_+.\ee The
corresponding relaxation time can be determined exactly by
realizing that the spectrum of $\cal L$ equals that of
$-(H\otimes\one+\one\otimes H)/2$ so that $\tau=1$ \footnote{That
the first part (denote it by $\cal F$) of the Liouvillian in
Eq.(\ref{eq:graphL}) does not contribute to the spectrum can be
seen by showing that $\tr{[(t{\cal F}+({\cal L-F}))^n]}$ with
trace in Liouville space is independent of $t$ for all
$n\in\mathbb{N}$. This can in turn be seen by observing that all
the contributions which are not powers of ${\cal L-F}$ are
nilpotent. As $\tr{[(t{\cal F}+({\cal L-F}))^n]}$ determines the
spectrum of $\cal L$ we can thus set $t=0$ without changing the
spectrum.}.

Let us now discuss the second type of commuting
Hamiltonians---those for which Eq.(\ref{eq:comU}) and $q<1$ are
incompatible. For this class we can still prove fast convergence
by making explicit use of the fact that frustration-free ground
states of commuting Hamiltonians have an efficient PEPS
representation. That is, when expanded in computational product
basis, the coefficients are given by a tensor-network whose
geometry resembles the lattice structure of the interactions. A
generic property of PEPS is \emph{injectivity} \cite{injectivity}
of local regions which is, in fact, a sufficient condition for the
state to be the unique ground state of its parent Hamiltonian.
Consider cases of commuting Hamiltonians for which the ground
state has this property. To specify the cp-map in Eq.(\ref{eq:T})
we need to sort the regions of interactions
$\lambda_1,\lambda_2,\ldots$ such that the union
$\Lambda_k=\bigcup_{i=1}^k\lambda_i$ has an intersection with
$\lambda_{k+1}$ but does not entirely cover it, i.e.,
$I_{k+1}:=\lambda_{k+1}\backslash \Lambda_{k}\neq\varnothing$.
Such an ordering is always achievable by possibly regrouping the
interactions into slightly larger regions. The reason for this
ordering is that any $U_{\lambda_k}$ which acts only on $I_k$,
will not alter the energies in all regions $\lambda_i$ with $i<k$.
That is, we have a weakened version of Eq.(\ref{eq:comU}).
Injectivity of the regions $I_k$ in the PEPS representation
implies then that there is always a unitary $U_{\lambda_k}$ (a
depolarizing set would work as well) such that $q<1$
\footnote{More precisely there exists a unitary $U$ on $I_{k}$
such that if $|\Phi\rangle$  has zero energy on $\Lambda_{k-1}$,
the rotated state $U|\Phi\rangle$ has non-zero overlap with the
zero-energy subspace corresponding to $\Lambda_{k}$. We argue by
contradiction. Assume that for all $U$ on $I_{k}$ this overlap and
therefore the one of the depolarized state would be zero. Every
normalized vector in the support of the depolarized state has the
form
$$|\phi\rangle =\sum_{\alpha,\beta}
\tr[A^\alpha X^\beta]|\alpha\rangle\ket{\varphi}\ket{\beta},$$
where $\alpha,\beta$ are product bases for $\Lambda_{k-1}$ and the
complement of $\Lambda_k$ respectively, $\varphi$ is a vector in
the Hilbert space of $I_k$ and $\tr$ denotes the contraction of
the tensors $A$ and $X$. The tensors $A$ are already the one of
the target state. Incorporating $U$ into $\varphi$, zero-overlap
would mean that $\langle\phi'|\phi\rangle=0$ for all
$$|\phi'\rangle =\sum_{\alpha,j,\beta} \tr[A^\alpha B^j
Y^\beta]|\alpha\rangle\ket{j}\ket{\beta},$$ with $B^j$ being the
tensor corresponding to region $I_k$ in the target state and $Y$
arbitrary (in particular $Y=X$). By the definition of injectivity
\cite{injectivity} there is, however, always a constant $c$ and a
vector $\varphi$ such that $c^{-1}\sum_j \langle j|\varphi\rangle
B^j=\one$ and thus $\langle\phi'|\phi\rangle=c$.} and we can thus
follow the above lines of argumentation. The only difference is
that due to the weakening of Eq.(\ref{eq:comU}) the energy does
not decrease homogeneously, but a low-energy region will grow
stochastically according to the $\Lambda_k$'s which requires extra
time proportional to the systems size, so that $\tau={\cal
O}(N\log N)$.

There are frustration-free ground states which belong to the
second class of commuting Hamiltonians but for which injectivity
does not hold (e.g. due to a degenerate ground state space). A
paradigmatic example is Kitaev's toric code state \cite{toric}
where one has a four-fold degeneracy. Due to lack of injectivity,
we have to prove $q<1$ separately which is, however, trivial in
this case since it is a stabilizer state so that $q=0$. The action
of the cp-map (or respective Liouvillian) can be understood as
moving all the excitations towards a single point where they can
mutually annihilate.

We turn now to another family of ground states of frustration free
Hamiltonians, namely MPS \cite{MPS}. Clearly, one possible
efficient way of preparing them using dissipation is to exploit
the fact that they can be obtained via a sequential application of
quantum gates \cite{sequential} together with the above DQC
scheme. In the following we will, however, focus on a different
way which does not require an additional time register.

For the sake of clearness, we will consider translationally
invariant Hamiltonians, although the analysis can be
straightforwardly extended to systems without that symmetry. We
will specify a cp-map to prepare states of the form
 \be
 \label{MPS}
 |\Psi\rangle = \sum_{i=1}^d {\rm tr}(A_{i_1}... A_{i_N}) |i_1\ldots
 i_N\rangle
 \ee
where, the $A$'s are $D\times D$ matrices. As before, we assume
the injectivity property which implies that $\Psi$ is the unique
ground state of a nearest neighbor frustration free 'parent'
Hamiltonian which has a gap. Denoting by $\rho$ the reduced
density operator corresponding to particles $k$ and $k+1$, $H_k$
and $P_k=1-H_k$ will denote the projectors onto its kernel and
range, respectively. Note that ${\rm tr}(P_k)=D^2$. We take
$N=2^n$ for simplicity, but this is clearly not necessary. We
construct the channel ${\cal T}$ in several steps. We first define
a channel acting on two neighboring particles $k, k+1$, as follows
 \[
 \RT_{r,c}(X):=P_{k}X P_{k} + \frac{P_{k}}{D^2}
 {\rm tr}(H_{k} X ).
 \]
Here, $k=2^{r-1}(2c-1)$ where $r=1,\ldots,n$ and
$c=1,\ldots,2^{n-r}$. The action of these maps has a tree
structure, where the index $r$ indicates the row in the tree,
whereas $c$ does it for the column. Now we define recursively,
 \be
 \label{channelS}
 \ST_{r,c}:= \frac{(1-\epsilon_r)}{2} (\ST_{r-1,2c} +
 \ST_{r-1,2c+1}) + \epsilon_r \RT_{r,c}.
 \ee
Here, $r=2,\ldots,n$, $c=1,\ldots,2^{n-r}$,
$\ST_{1,c}:=\RT_{1,c}$, and $\epsilon_{r+1}=1/M^r$ where $M=CN^2$
and $C\gg 1$ (see appendix). Note that $\ST_{r,1}$ acts on the
first $2^r$ particles, $\ST_{r,2}$ on the next $2^r$, and so on.
We finally define
 \be
 \label{cpmapT}
 \TT:= (1-\epsilon_{n+1}) \ST_{n,1} +
 \epsilon_{n+1} \RT_{n,2}.
 \ee
In the appendix we show that this map achieves the fixed point (up
to an exponentially small error in $C$) in a time ${\cal
O}(N^{\log_2(N)})$. The intuition behind the cp-map (\ref{cpmapT})
is that the channels ${\cal S}_{1,c}$, which are the ones that
most often applied, project the state of every second nearest
neighbors onto the right subspace. Then ${\cal S}_{2,c}$ do the
same with half of the pairs which have not been projected. Then
${\cal S}_{3,c}$ does the same on half of the rest, and so
on.\vspace*{6pt}

In conclusion, we have investigated the computational power of
purely dissipative processes, and proven that it is equivalent to
that of the quantum circuit model of quantum computation. We have
also shown that dissipative dynamics can be used to create ground
states (like MPS or PEPS) of frustration free Hamiltonians of
strongly correlated quantum spin systems. This implies the
existence of dissipatively driven quantum phase transitions,
something which could be experimentally tested using atoms or
ions.

Let us stress that we have been concerned here with a
proof--of--principle demonstration that dissipation provides us
with an alternative way of carrying out quantum computations or
state engineering. We believe, however, that much more efficient
and practical schemes can be developed and adapted to specific
implementations. We also think that these results open up some
interesting questions which deserve further investigation. For
example, how the use of fault tolerant computations can make our
scheme more robust. Or how one can design translationally
invariant cp-maps that prepare MPS more efficiently. Or the
importance and generality of the set of commuting Hamiltonians,
which we believe to be intimately connected to the fixed points of
the renormalization group transformations on PEPS (as it happens
with MPS \cite{renormalization}). Furthermore, the model of DQC
might well lead to the construction of new quantum algorithms, as
e.g. quantum random walks can more easily be formulated within
this context. Finally, other ideas related to this work can be
easily addressed using the methods introduced, e.g., thermal
states of commuting Hamiltonians can be engineered using DSE since
the Metropolis way of sampling over classical spin configurations
can be adopted to the case of commuting operators. Similar
techniques could be applied to free fermionic and bosonic systems,
and, more generally, it should be possible to device DSE-schemes
converging to the ground- or thermal states of frustrated
Hamiltonians by combining unitary and dissipative dynamics.

\section*{Acknowledgments}
We thank D. Perez-Garcia for discussions and acknowledge financial
support by the EU project SCALA, the DFG, Forschungsgruppe 635,
and the Munich Center for Advanced Photonics (MAP).

\section{Appendix: Efficiency of DQC}

In this section we will prove that the Liouvillian defined by the
Lindblad operators
\bea L_i&=&|0\rangle_i\langle 1|\otimes |0\rangle_t\langle 0|\\
L_t&=& U_t\otimes|t+1\rangle\langle
t|+U_t^\dagger\otimes|t\rangle\langle t+1|\eea is gapped. More
specifically, it holds that for any initial condition $\rho(0)$,
we can show that $\|\rho(t)-\rho(\infty)\|\leq \epsilon$ in a time
$t$ that scales logarithmically in $1/\epsilon$ and quadratically
in $1/T$.

The Liouvillian is defined as the (non-symmetric) matrix
\[\mathcal{L}=\sum_{\alpha}L_\alpha\otimes\bar{L}_\alpha-\frac{1}{2}\left(\sum_\alpha
L_\alpha^\dagger L_\alpha\otimes I+I\otimes \sum_\alpha
L_\alpha^\dagger L_\alpha\right)\] where $\alpha$ runs over the
labels of all Lindblad operators. We will bring this Liouvillian
into a simpler form by doing two eigenvalue-preserving similarity
transformations. First, we apply the unitary

\[W=\sum_t U_tU_{t-1}...U_1\otimes |t\rangle\langle t|\]

and observe that the spectrum of the Liouvillian is the same as if
the quantum circuit would only have consisted of identity gates. The
spectrum of the Liouvillian is hence independent of the actual
computation that we want to do (note that this was also the case in
the context of adiabatic quantum computation). Without loss of
generality, we therefore assume $\forall t: U_t=I$. Second, we
diagonalize the part acting on the logical qubits by the similarity
transformation $\mathcal{L}'=X\mathcal{L}X^{-1}$ where

\bea X&=&\left(X_1\otimes X_2\otimes ...X_N\right)\otimes
|00\rangle_t\langle 00|+I\otimes \left(I-|00\rangle_t\langle
00|\right)\nonumber\\
X_i&=&|00\rangle\langle 00|+|00\rangle\langle
11|-|11\rangle\langle 11|+|01\rangle\langle 01|+|10\rangle\langle
10|. \nonumber\eea

Note that the double indices arising in those expressions reflect
the fact that the Liouvillian acts on the tensor product of the
physical space with itself. The part acting on the logical qubits
is now completely diagonal, and $\mathcal{L}'$ becomes

\begin{eqnarray*} \mathcal{L}'&=&\sum_{i_1,i_2,...i_N,j_1,...j_N=0}^{1}
|i_1...j_1...\rangle\langle i_1...j_1...|\otimes \tilde{L}_{ij}\\
\tilde{L}_{ij}&=&\tilde{L}+\lambda_i |0\rangle_t\langle 0|\otimes
I_t+\lambda_j
I_t\otimes |0\rangle_t\langle 0|\\
\lambda_i&\equiv&\sum_{k=1}^N i_k\\
\tilde{L}&=&\sum_t \tilde{L}_t\otimes
\tilde{L}_t-\frac{1}{2}\left(\sum_t \tilde{L}_t^\dagger
\tilde{L}_t\otimes
I_t+I_t\otimes \sum_t \tilde{L}_t^\dagger \tilde{L}_t\right)\\
\tilde{L}_t&=&|t\rangle\langle t+1|+|t+1\rangle\langle t|
\end{eqnarray*}

The problem of calculating the eigenvalues of $L$ is therefore
reduced to the problem of calculating the eigenvalues of the
$T^2\times T^2$ matrices $\tilde{L}_{ij}$ for all possible
positive integer numbers $\lambda_i,\lambda_j$. It happens that
$L_{ij}$ is block-diagonal, and consists entirely of diagonal
elements of the form $(-2)$, $(-1-\lambda_i)$, $(-1-\lambda_j)$,
of $2\times 2$ blocks of the form \[\left(
         \begin{array}{cc}
           -1-\lambda_i & 1 \\
           1 & -1-\lambda_j \\
         \end{array}
       \right) ,\left(
                  \begin{array}{cc}
                    -2 & 1 \\
                    2 & -2 \\
                  \end{array}
                \right)
\]
and of tridiagonal matrices of the form

\begin{eqnarray*}-&(1+\lambda_i+\lambda_j)|0\rangle\langle
0|-2\sum_{t=2}^{T-1}|t\rangle\langle t|-|T\rangle\langle T|\\
&\hspace{2cm}+\sum_{t=1}^{T-1} \left(|t\rangle\langle
t+1|+|t+1\rangle\langle t|\right)\end{eqnarray*}

The first classes of diagonal blocks have all eigenvalues strictly
smaller than $-1$, and therefore correspond to terms that converge
extremely fast. The gap is determined by the last class of
tridiagonal matrices, and that one corresponds exactly to the
class of matrices that has been extensively studied in the context
of random walks. The only matrix with eigenvalue $0$ is the one
with $\lambda_i=\lambda_j=0$, which proves that the fixed point is
unique. The gap of $L_{00}$ can be calculated exactly, and is
given by
\[2\left(\cos\left(\frac{\pi}{T+1}\right)-1\right)\simeq -\frac{\pi^2}{(T+1)^2}.\]
The largest eigenvalue for the cases $(ij)\neq (00)$ is obtained
for $\lambda_i+\lambda_j=1$, and is given by
\[2\left(\cos\left(\frac{\pi}{2T+3}\right)-1\right)\simeq -\frac{\pi^2}{(2T+3)^2}.\]
It follows that the gap of the Liouvillian is larger than $1/T^2$,
which we set out to prove. In principle, this does not suffice to
prove that convergence will be reached at a time in the order of
$1/T^2$, as the Liouvillian might have exponentially large Jordan
blocks; it can readily be checked that this is not the case here.

Let us next check what happens if we use Kitaev's unary encoding
\cite{KitaevQMA} of the time register such as to make all Lindblad
operators strictly local. We will have to replace $L_t$ by
\[L_t=I\otimes |1\rangle\langle 1|\otimes X \otimes |0\rangle\langle
0|\otimes I\] where $X$ is acting on the $t$'th qubit in the unary
encoding, and add Lindblad terms that force the system to converge
into the allowed subspace for the time Hamiltonians:
\[L_q=I\otimes |0\rangle\langle 0|\otimes |0\rangle\langle 1|\otimes
I.\] Furthermore, we have to replace
\[L_i= (I\otimes |0\rangle_i\langle 1|\otimes I)\otimes (|0\rangle\langle
0|\otimes I).\]

It now happens that the relevant spectrum of the new corresponding
Liouvillian is unchanged: all the terms $L_i,L_t,L_q$ are such
that they map density operators that are in the right subspace of
the unary encoding into the right subspace, and furthermore the
only terms that connect the right with the wrong subspace are
$L_q$ which can map wrong states right ones. The Liouvillian is
therefore block-upper-diagonal, and the eigenvalues are completely
determined by the diagonal blocks. All the eigenvalues in the
wrong block turn out to be smaller than $-1$, and therefore the
relevant eigenvalue lie in the right block. The gap is therefore
left unchanged.

Making use of the recent results of Oliveira and Terhal
\cite{Oliveira}, and of Aharonov, Gottesman and Kempe \cite{AGK},
it can also be shown that the same computational power is retained
if the Lindblad operators are only acting on nearest neighbour
qubits in a 2-D square lattice or on nearest neighbours in a
1-dimensional spin chain of 12- level systems.

Let us conclude this section by explaining why the gap will
considerably change if we were to use the master equation approach
to solve general problems in the class NP. The idea would be to
put a penalty term on one of the output qubits corresponding to
getting the right answer, and relaxing the constraints on some
input bits. First note that the construction of the Feynman
Hamiltonian \cite{Feynman} is only possible when all quantum gates
are unitary; hence the whole circuit is reversible. This implies
that the problem only makes sense if some input qubits to the
quantum circuit are initialized, as otherwise any output can be
obtained. Because of the fact that both input and output qubits
must be \emph{initialized}, we cannot replace the actual gates
with the identity gate, and as a result the Liouvillian does not
have a nice block-diagonal structure anymore with only blocks of
polynomial size, but we get exponentially large blocks. Such
blocks will typically lead to exponentially small gaps.

\section{Appendix: Engineering dissipation}

Here we show how to engineer the local dissipation which gives rise
to the master equations (\ref{Lindblad}) and cp--maps (\ref{eq:T}).
They are composed of local terms, involving few particles (typically
two), so that we just have to show how to implement those. In order
to simplify the exposition, we will treat those particles as a
single one and assume that one has full control over its dynamics
(e.g. one can apply arbitrary gates).

Let us start with the cp-maps. It is clear that by applying a
quantum gate to the particle and a 'fresh' ancilla and then tracing
the ancilla one can generate any physical action (i.e. cp-map) on
the system. Furthermore, by repeating the same process with short
time intervals one can subject the system to an arbitrary time
independent master equation. This last process may not be efficient.
An alternative way works as follows. Let us assume that the ancilla
is a qubit interacting with a reservoir such that it fulfills a
master equation with Liouville operator $L_a=\sqrt{\Gamma}\sigma_-$,
where $\sigma_-= |0\rangle\langle 1|$. Now, we couple the ancilla to
the system with a Hamiltonian $H=\Omega (\sigma_-L^\dagger+
\sigma_-^\dagger L)$. In the limit $\Gamma\gg \Omega$ one can
adiabatically eliminate the level $|1\rangle$ of the ancilla
\cite{Cohen} by applying second order perturbation theory to the
Liouvillian. This is done as follows: the unperturbed Liouvillian
can be written as $\mathcal{L}_0=QDQ^{-1}$ with $Q=Q^{-1}$ the
eigenvectors of $\mathcal{L}_0$; writing the perturbed eigenvectors
as $Q\exp(\Omega/\sqrt{\Gamma} X)$, we solve the equation
$-XD+DX+Q\mathcal{L}_HQ=0$ with $\mathcal{L}_H$ the perturbation
arising from the Hamiltonian part; the effective Liouvillian is then
given by $-\Omega^2/\Gamma XDX$. In this way we obtain a master
equation for $\rho$ describing the system alone, with Liouville
opeartor $\Omega/\sqrt{\Gamma} L$. By using several ancillas with
Hamiltonians $H=\Omega (\sigma_-L_i+ \sigma_-^\dagger L_i^\dagger)$
and following the same procedure we obtain the desired master
equation. Although we have not specified here a physical system, one
could use atoms. In that case, the ancilla could be an atom itself
with $|0\rangle$ and $|1\rangle$ an electronic ground and excited
level, respectively, so that spontanous emission gives rise to the
dissipation. The coupling to the system (other atoms) could be
achieved using standard ideas used in the implementation of quantum
computation using those systems \cite{PhysToday}.

\section{Appendix: Efficiency of DSE for matrix product states}

We show that the cp-map defined in the text for the creation of
MPS converges in sub--exponential time.

Let us denote by $\HH_r$ the ground subspace of the Hamiltonian
$\sum_{k=1}^{r-1} H_k$ corresponding to the zero eigenvalue, and
by $q_n$ the projector onto the orthogonal subspace. The basic
idea is that by applying the map $\ST_{r,1}$ a sufficiently large
number of times, $L_r$, to any density operator, $\rho$, we obtain
a state which is practically supported on $\HH_r$. We will show
that the error $\mu_r^{L} :={\rm tr}(q_r \ST_{r,1}^L(\rho)$ can be
made arbitrarily small. In particular, for $n=\log_2 N$ we just
have to choose $L={\cal O}(N^{\log_2(N)})$.

In order to simplify the discussion, we will take into account the
following: $\ST_{r+1,1}^L$ with $M:=L\epsilon_{r+1}\gg 1$ will be
a sum of contributions which will typically have the form
 \be
 \label{sequence}
 \ST_{r,1}^{L_1} \ST_{r,1}^{L'_1}\RT_{r+1,1} \ST_{r,1}^{L_2}\ST_{r,1}^{L'_2}
 \RT_{r+1,1} \ldots
 \ee
where $L_i\sim (1-\epsilon_{r+1})/2\epsilon_{r+1}$ and where the
channel $\RT_{r,1}$ appears $\sim L\epsilon_{r+1}$ times. We
define: $L_r=1/2\epsilon_{r+1}$; ${\cal s}_r=
\ST_{r,1}^{L_r}\ST_{r,2}^{L_r}$; ${\cal r}_{r+1}=\RT_{r+1,1} {\cal
s}_r$; $\rho_l={\cal s}_r {\cal r}_{r+1}^{l-1}$;
$\rho_l'=\RT_{r+1,1}(\rho_l)$; $X_l={\rm tr}(p_{r+1}\rho_l)$;
$Y_l={\rm tr}(p_{r+1}\rho'_l)$. We will approximate
$\ST_{r+1,1}^L\simeq \tilde \RT^{L\epsilon_{r+1}}_{r+1}$. Using
the fact that for $k<2^r$, $P_k p_r=p_rP_k=p_r$ we have that $
{\rm tr}[p_{r+1,1} \ST_{r,i}(\sigma)] \ge {\rm tr}[p_{r+1,1}
\sigma]$ for $i=1,2$, and thus $X_{l}\ge Y_{l-1}$. It is easy to
show using  the same thing that
 \be
 Y_l=X_l + \frac{1}{D^2} {\rm tr} [p_{r+1} {\rm tr}_0(Q\rho_l
 Q)].
 \ee
Here we have simplified the notation: tr$_0$ denotes the trace
with respect to the particles $k_0,k_0+1$, and $Q=Q_{k_0,k_0+1}$.

Now, using that for any projectors $P$ and $Q=\one-P$,
$(\sqrt{\epsilon} P\pm Q/\sqrt{\epsilon})\rho (\sqrt{\epsilon}
P\pm Q/\sqrt{\epsilon})\ge 0$ it follows that $\rho_l\ge p_r
\rho_l p_r - 3\sqrt{\mu_r^{L_r}}$. Finally, using the properties
of MPS \cite{MPS} it is easy to show that  ${\rm tr} [p_{r+1}{\rm
tr}_0(Qp_r\rho_lp_r Q)]/D^2$ is lower bounded by a $z
(1-\mu_r^{L_r}) (1-X_l)$, where $z$ is a constant. Thus we obtain
 \be
 X_{l+1} \ge Y_l\ge X_{l}- 3 \sqrt{\mu_r^{L_r}} + z (1-\mu_r^{L_r})(1-X_l).
 \ee
Iterating this expression we obtain
 \be
 \mu_{r+1}^{L_r+1} \le \frac{3}{z
 (1-\mu_r^{L_r})}\sqrt{\mu_r^{L_r}}+
 [1-z(1-\mu_r^{L_r})]^{L_{r+1}\epsilon_{r+1}}.
 \ee
After some lengthy algebra, we obtain that if we choose
$\epsilon_r/\epsilon_{r+1}=M=CN^2$, and $L_r=1/\epsilon_{r+1}$ the
final error will be $(3/z)^2 (1-z)^C$ whereas the number of
applications of the map $L_{\log_2N}=N^{2\log_2N+\log_2C}$. Thus,
by choosing $C$ sufficiently large we can always make the error
arbitrarly small with a subexponential number of applications of
the map.


\begin{thebibliography}{99}
\bibitem{note1}
This mechanism is different to the one usually considered in
condensed matter physics, where instead of the system's steady
state, the ground state of the system--plus--environment
experiences abrupt changes when the interaction is modified
\cite{referenceCMP}.
\bibitem{note3}To simplify matters we assume that the number of
interactions is proportional to the number of sites, and denote
it, with some abuse of notation, by $N$ as well.
\bibitem{note4}
Note that the reduction in the time of convergence comes from the
fact that the cp-map $\cal T$ applies one operation with
probability $p_\lambda$ at a time, whereas in $\cal L$ all sites
couple to their respective baths simultaneously.
\end{thebibliography}
\end{document}